\documentclass[a4paper]{article}
\usepackage{epsfig,amsmath,amsfonts,amssymb,setspace,multirow,textcomp}
\usepackage[T1]{fontenc}
\makeatletter
\renewcommand{\@oddhead}{\textit{Advances in Astronomy and Space Physics} \hfil}
\renewcommand{\@evenfoot}{\hfil \thepage \hfil}
\renewcommand{\@oddfoot}{\hfil \thepage \hfil}
\makeatother

\renewenvironment{thebibliography}[1]{\begin{oldthebibliography}{#1}\setlength{\parskip}{0ex}\setlength{\itemsep}{0ex}}{\end{oldthebibliography}}

\begin{document}
\fontsize{11}{11}\selectfont 
\title{THE STRUCTURE OF THE TEST FUNCTION FOR PHENOMENOLOGICAL MODELLING OF ECLIPSING BINARIES}
\author{\textsl{M.\,G.~Tkachenko}}
\date{\vspace*{-6ex}}
\maketitle
\begin{center}{\small Department "Mathematics, Physics and Astronomy",\\ 
Odessa National Maritime University, Mechnikova Str., 34,\\ 
65029, Odessa, Ukraine\\
{\tt masha.vodn@yandex.ua}}
\end{center}

\begin{abstract}
The dependence of the test function on the phenomenological parameters used in the "NAV" ("New Algol Variable") algorithm (Andronov, 2012Ap.....55..536A) is studied. Due to a presence of local minima, the method of minimization contains two steps: the "brute force" minimization at a grid in the 4D parameter space, and further iterations using the differential corrections. This method represents an effective approximation of the light curve using the special pattern (shape) separately for the primary and secondary minima. The application of the method to concrete stars is briefly reviewed.
\\[1ex]
\end{abstract}

\section*{\sc introduction}
\indent \indent Currently nearly 400 000 variable stars are listed in the Variable Stars Index (VSX, {\it http://aavso.org/vsx}). In the General Catalogue of Variable Stars (Samus, et al., 2007-2016 \cite{samus}), the on-line version of which is available at {\it http://www.sai.msu.su/gcvs/gcvs/}, there are currently 52011 objects with official GCVS names. Among them, there are 10845 objects classified as the eclipsing ones, distributed among the subtypes as 5294 (EA), 3018 (EB), 1434 (EW), 1099 (E). Only few dozens of these objects were studied carefully, using not only photometric, but also spectral and (rarely) polarimetrical observations. For such rare objects, the "physical" modelling is possible with a determination of radii, masses, temperatures. The "standard" approach is so-called "Wilson -- Devinney" model \cite{wd71}, \cite{wd94}, which was realized in some famous programs "Binary Maker" \cite{bm}, Phoebe \cite{phoebe} and in the set of programs elaborated by S.Zo{\l}a et al. \cite{zola1}, \cite{zola2}. Different problems of the physical modelling were described e.g. by \cite{cher1993}, \cite{kopal1959}, \cite{kallrath2009}.

For the majority of stars, there are only photometric observations, often obtained with one (or no) filter, thus the "physical" modelling is not possible because of unknown temperatures of the components and their mass ratio. In this case, only "phenomenological" modelling is available, which characterizes a smaller number of parameters, which describe the light curve, namely, the period $P,$ the initial epoch $T_0,$  brightness at primary maximum $m_{max}$ and primary minimum $m_{min},$ the duration of the eclipse $D.$ Additionally in the section "remarks" in the GCVS, are listed the brightness at the secondary minimum and (if different from $m_{max}$) at the secondary maximum and the phase shift of the secondary minimum in respect to the phase 0.5 (significant in a case of elliptical orbits) \cite{samus}, \cite{tsessevich}. 

Typically the values of brightness and phases are being determined using local approximations of observations in intervals, which include the extrema (either maximum, or minimum) (e.g. \cite{andr2005}).
 
Some methods used the trigonometrical polynomial approximation of the complete light curve \cite{ruc1993}. Also there is a set of studies based on the "simplified physical" model, which suggests spherically symmetrical components with uniform brightness distribution \cite{shulberg},\cite{tk2013},\cite{malkov2007}. 

And recently were actively used more accurate methods, based on special shapes (patterns) of the minima \cite{andr2010},\cite{andr2012}, \cite{mikulas2015}.   

Another algorithm to determine the statistically optimal degree $s$ of the trigonometrical polynomial (sometimes called the "restricted Fourier series") is based on the minimization of the r.m.s. estimate of the accuracy of the smoothing function at the arguments of observations \cite{a1994}, \cite{andr2003}. This method was effectively applied also for pulsating Mira-type variables \cite{kuda1996}.

In this paper, we compare previously used approximations with that using the special shape (pattern) and study behavior of the test function in the parameter space. For illustration, we have used $n=1000$ values of the phenomenological "NAV" function {\cite{andr2012}) with fixed parameters, which is a model for the light curve of an EA-type eclipsing binary.

\section*{\sc the methods of calculations: trigonometric polynomial}

There were oversimplified models, which could be effective for automatic classification of numerous newly discovered variable stars using the surveys, e.g. the "EA" catcher with a parabolic shape of minima of equal width and different depth \cite{andr2000}. 

More recently, Papageorgiou et al. \cite{papa2014} proposed a model of parabolic shape either for the out-of-the eclipse parts (phases (0.1 -- 0.4) and 0.6 -- 0.9) of the light curve), or to the eclipses (fixed phases (-0.2 -- 0.2), (0.3 -- 0.7)). The correspoding curve is shown in  Fig.~\ref{fig1}. One may note a reasonably good appoximation out of eclipses, but a bad approximation at the phases of minima because of an overestimated eclipse width. 

\begin{figure}[!h]
\centering
\epsfig{file = 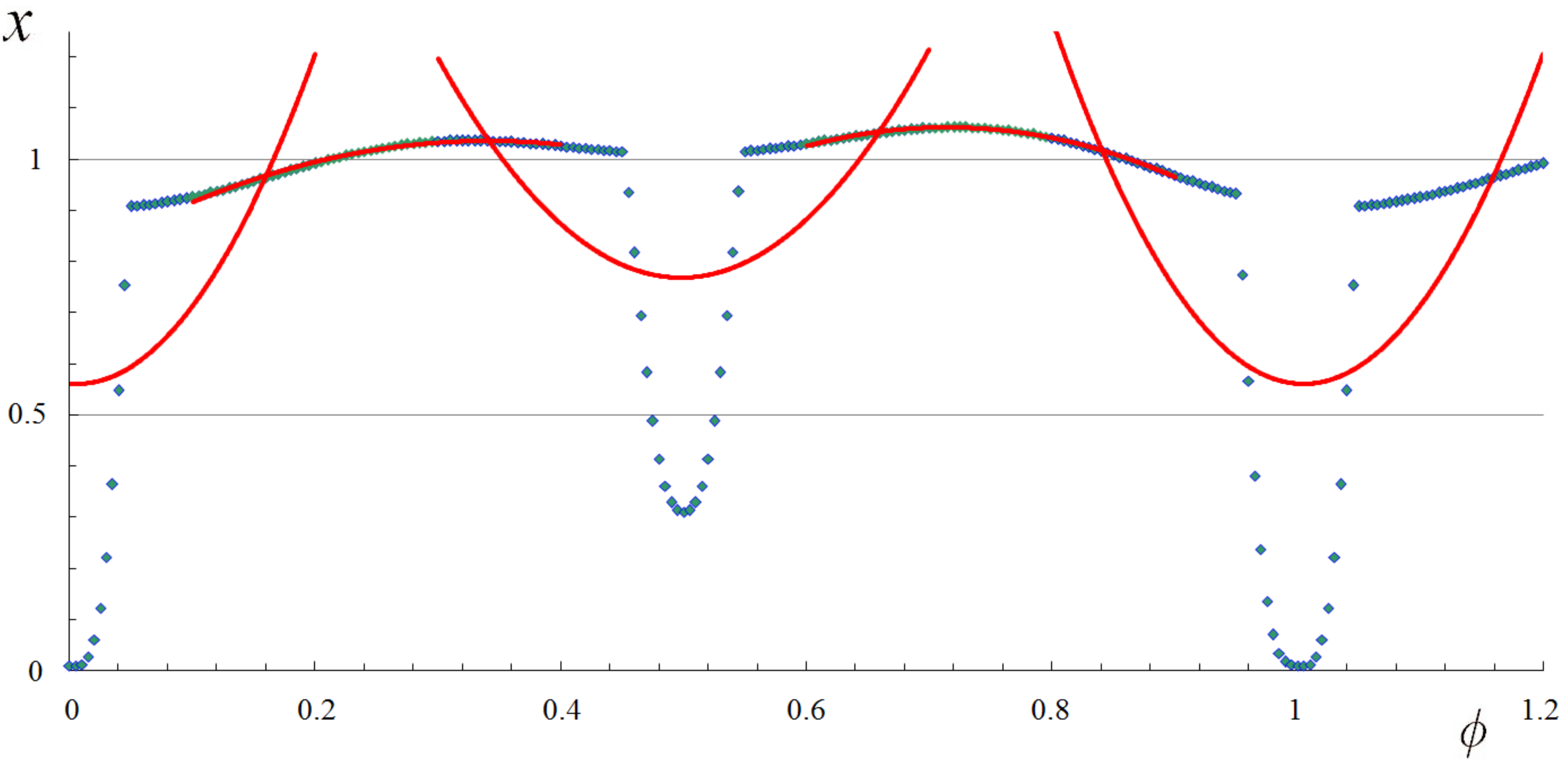,width = 0.5\linewidth}
\caption{The model light curve and its approximation by parabola at the intervals of phases centered on minima and maxima, as proposed by (Papageorgiou et al., 2014)}\label{fig1}
\end{figure}

\begin{figure}[!h]
\centering
\epsfig{file = 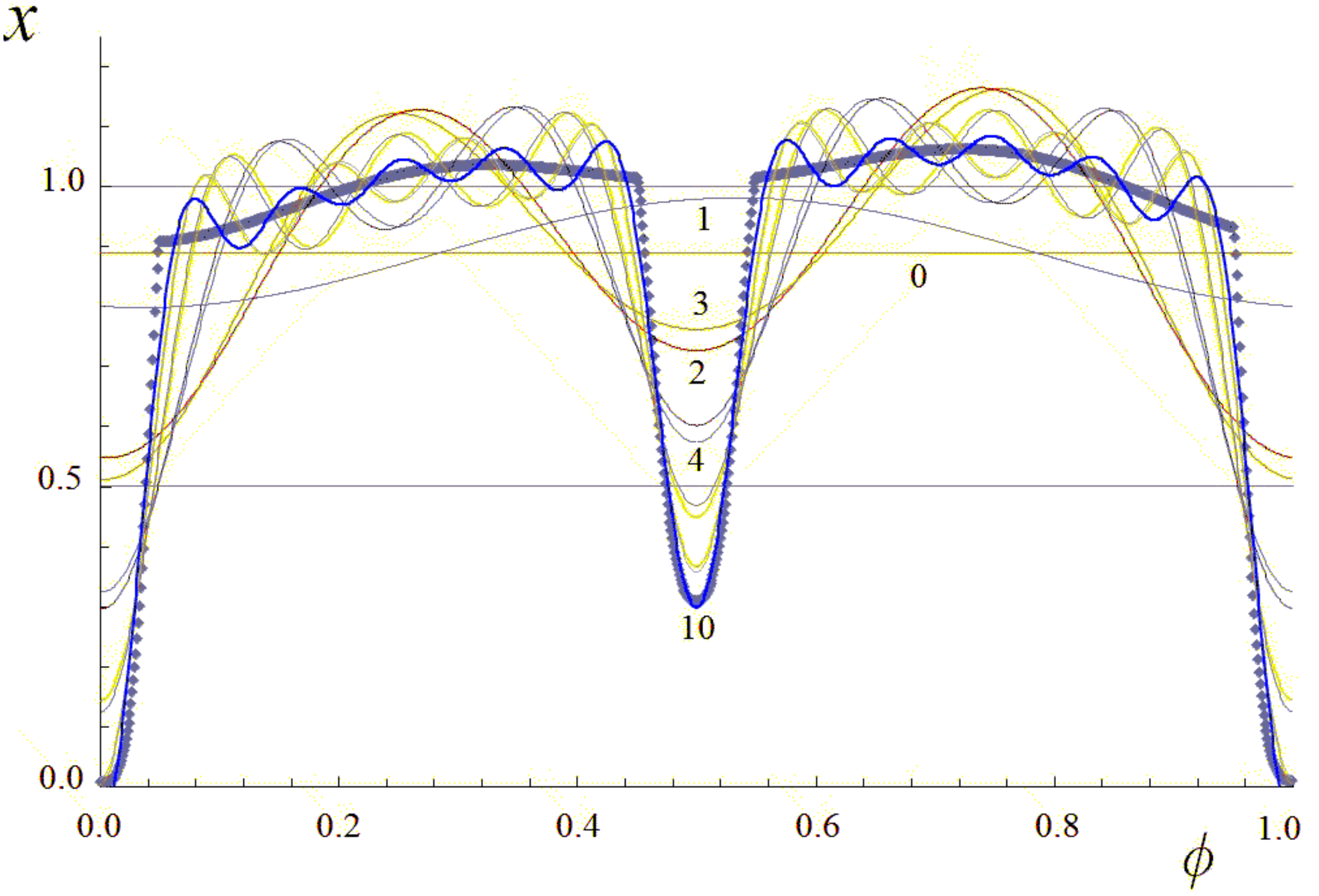,width = 0.5\linewidth}
\caption{Trigonometrical polynomial approximations of the phenomenological light curve. The degree s is shown by numbers near corresponding curves.}\label{fig2}
\end{figure}

Moreover, the light curve is not continuous. In the much earlier  "EA" catcher \cite{andr2000}, the smoothing function was continuous, and the width (as well as the phase shift) was determined using non-linear least squares fitting (see examples of this and other functions in \cite{andr2016a}, \cite{andr2016b}, \cite{andr2016c}). The approximation of the extrema of variable stars using the algebraic polynomial of the statistically optimal degree was realized in the software \cite{b2003}, \cite{andrandr2015}. The separate case of abrupt decline and inclined parts of the light curve, when the analytical function are obviously bad approximations, was discussed by \cite{andrandr2014}.
The fixed width in the method \cite{papa2014} leads to systematic differences between the observations and the approximation.

Next approximation is a trigonometrical polynomial 
\begin{eqnarray}
 x_c(\phi)&=& C_1+\sum_{j=1}^s(C_{2j}\cos(2\pi j\phi)+C_{2j+1}\sin(2\pi j\phi))\nonumber\\
&=&C_1+\sum_{j=1}^s R_j\cos(2\pi j(\phi-\phi_{0j})),
\label{eq1}
\end{eqnarray}
where $\phi$ is the phase, $\phi_{0j}$ are initial phases corresponding to the maximum of the wave with the $j^{th}$ term of the sum, and $R_j$ are corresponding semi-amplitudes.
The coefficients $C_\alpha$ $(\alpha=1..m=1+2s)$ are determined using the least squares method.

The approximations, which use the trigonometric polynomial of different degrees $s$, is shown in  Fig.~\ref{fig2}. One may note an expected refinement of the approximation with an increasing $s.$ The coefficients $C_{2j}$ are shown in Fig.~\ref{fig3}. They describe terms with a cosine function, so the "symmetrical" part of the light curve. For even $j,$ the absolute values are typically larger, what is explained by a similarity in depth of the primary and secondary minima, as the coefficients with even $j$ approximate a mean light curve with a double frequency, and the coefficients with odd $j$ approximate the difference:
\begin{eqnarray}
\frac{x_c(\phi)+x_c(\phi+0.5)}{2}&=& C_1+\sum_{k=1}^{s/2}(C_{4k}\cos(4\pi k\phi)+C_{4k+1}\sin(4\pi k\phi)),\\
\frac{x_c(\phi)-x_c(\phi+0.5)}{2}&=& \sum_{k=1}^{s/2}(C_{4k-2}\cos(2\pi (2k-1)\phi)+C_{4k-1}\sin(2\pi(2k-1)\phi)).\nonumber
\label{eq2}
\end{eqnarray}

\begin{figure}[!h]
\centering
\epsfig{file = 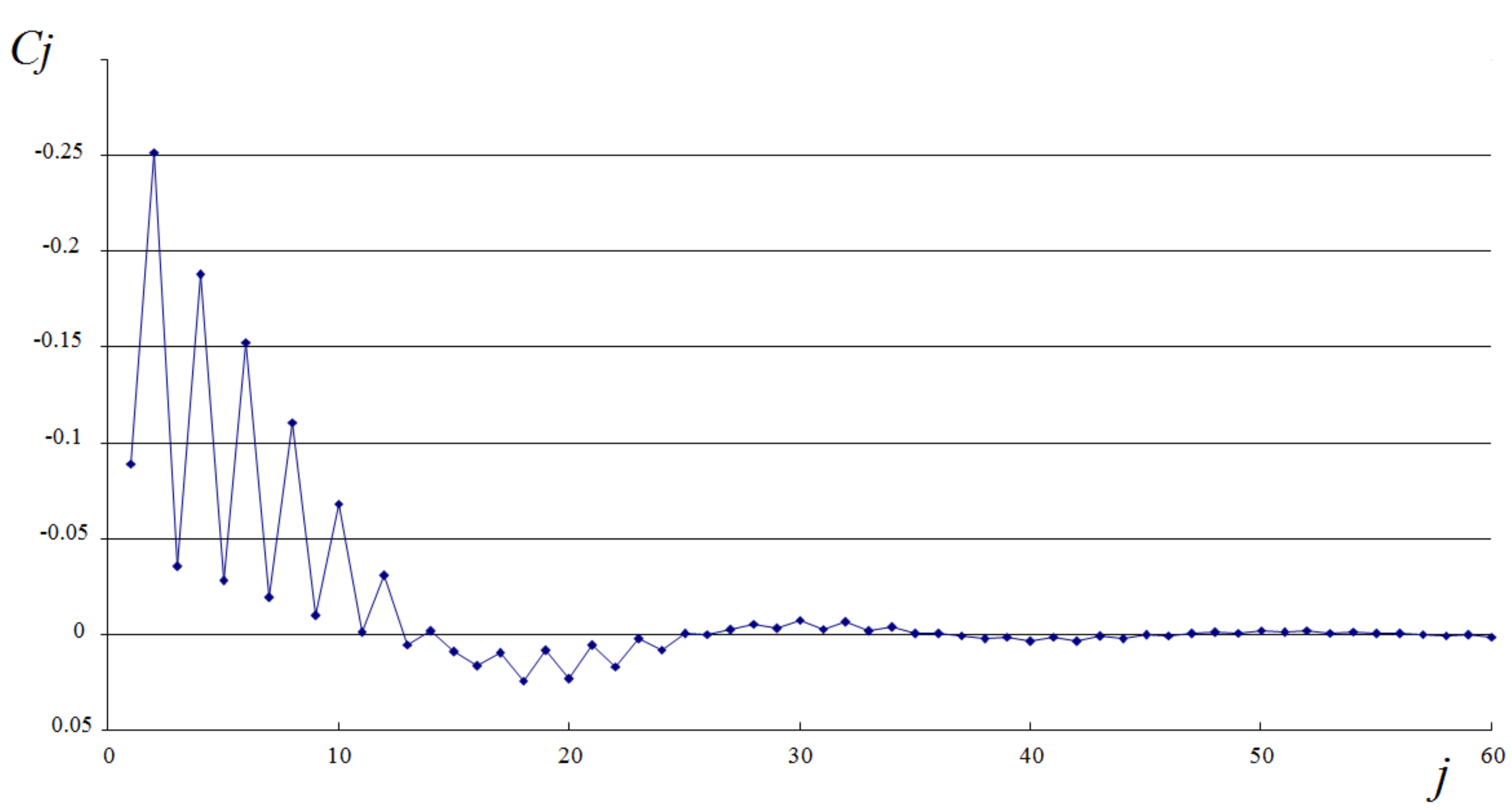,width = 0.5\linewidth}
\caption{Dependence of coefficients $C_j$ on $j$.}\label{fig3}
\end{figure}

\begin{figure}[!h]
\centering
\epsfig{file = 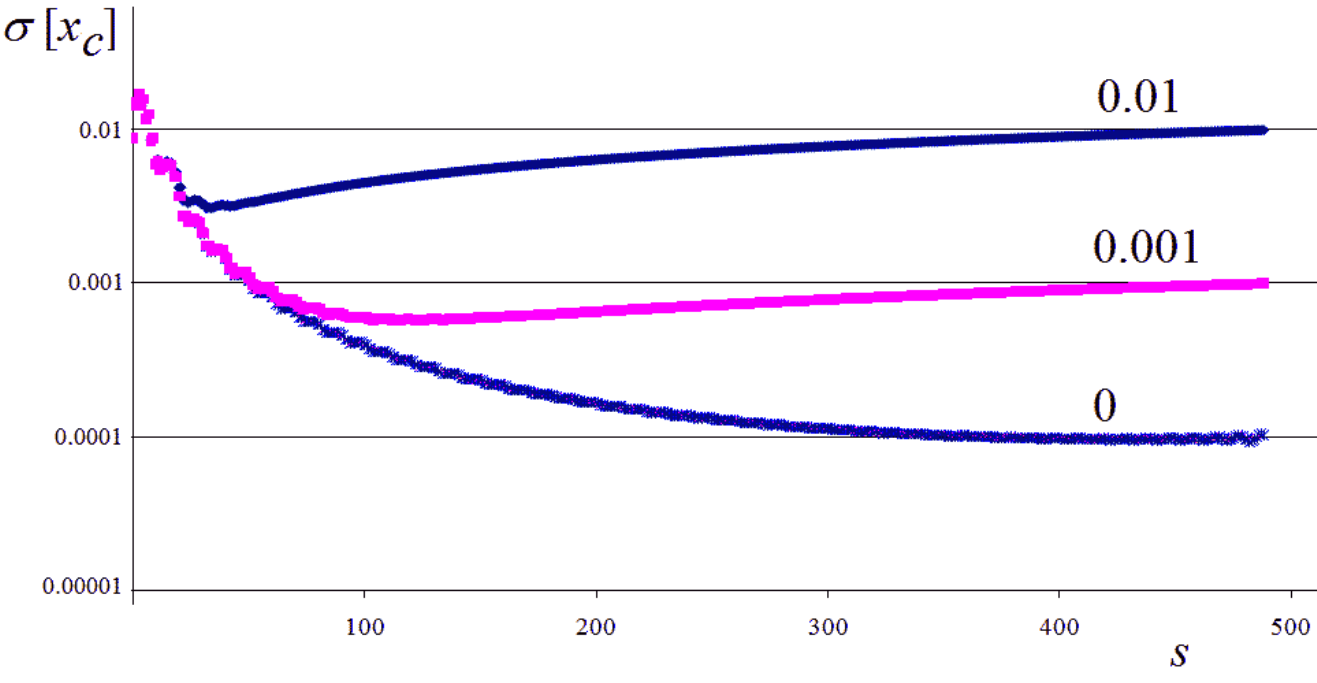,width = 0.5\linewidth}
\caption{Dependence of the mean squared error estimate $\sigma[x_c]$ of the approximation for an additional noise with a r.m.s value of 0  (bottom), 0.01 (up) , 0.001 (middle).}\label{fig4}
\end{figure}

Additionally, if the O'Connell effect is practically absent (what is the case for the majority of objects) \cite{papa2014}, the terms with sine vanish, and one gets only sums of terms with cosines. 

At this dependence, the coefficients tend to zero, but too slowly. E.g. the last coefficient exceeding an arbitrary limiting value of 0.001 occurs at $j=64,$ so the corresponding number of parameters $m=1+2\cdot 64=129$ is extremely large.

For the determination of the statistically optimal value of $s,$ different criteria may be used (e.g. \cite{a1994}, \cite{andr2003}). The first is based on the Fischer's criterion, which assumes uncorrelated observational errors obeying the normal distribution. For our data set (which contains computed values without any noise), this criterion is not applicable, as the deviations between the data and the approximation are systematic and not random. For real stars, we used this criterion as well (e.g. \cite{andr2016b},\cite{andr2003}).

The second criterion is based on minimization of the r.m.s. accuracy estimate $\sigma[x_c]$ of the approximation $x_c(\phi)$ at the arguments of observations $\phi_k$: 
\begin{eqnarray}
\sigma^2[x_c]&=&\frac{m}{n}\sigma^2_{0m},\nonumber\\
\sigma^2_{0m}&=&\frac{\Phi_m}{n-m},\\
\Phi_m&=&\sum_{k=1}^{n}w_k\cdot(x_k-x_c(\phi_k))^2.\nonumber
\label{eq3}
\end{eqnarray}
Here $\sigma_{0m}$ is a "unit weight error", $\Phi_m$ as a "test" ("target") function to be minimized in the parameter space.

The dependence of $\sigma[x_c]$ on the number of parameters $m$ is shown in Fig.~\ref{fig4}. In fact, it may be split into two almost monotonical sequencies for even and odd degrees of the trigonometric polynomial (as a consequence of the separate dependencies of the coefficients described above). The bottom sequence show a systematic decrease with $s,$ so formally the degree of the trigonometric polynomial should be extremely large, close to $n/2,$ i.e. the approximation tends to an interpolating function. This is because the data are precisely described by a function. 

In a real situation, the statistical errors are present, leading to qualitative and quantitative changes. For an illustration, we have suggested an additional observational noise with a standard error of (arbitrarily) 0.001 and 0.01. The resulting dependencies show a broad, but distinct minimum in Fig.~\ref{fig4} at $s=132$ and $s=34,$ respectively. 

The additional noise shifts the position of the minimum of the dependence of the r.m.s. value of the accuracy of the approximation towards smaller values, leading to the systematic shifts. Anyway, the degree is very large, leading to a significant number of coefficients, which are not statistically significant.

\section*{\sc the methods of calculations: the NAV algorithm}

To decrease the number of the parameters, Andronov \cite{andr2010}, \cite{andr2012} proposed the following approximation, which was called "the NAV" ("New Algol Variable") algorithm:
\begin{eqnarray}
 x_c(\phi)&=& G_1+G_2\cos(2\pi\phi)+G_3\sin(2\pi\phi)+\nonumber\\
&& +G_4\cos(4\pi\phi)+G_5\sin(4\pi\phi)+\\
&& +G_6H(\phi-C_4, C_1,C_2)+G_7H(\phi-C_4+0.5, C_1,C_3).\nonumber                                           \label{eq11}
\end{eqnarray}
where the shape (pattern) is localized to the phase interval
\begin{equation}
 H(\zeta,C_1,\beta)=\left\{
\begin{array}{ll}
V(z)=(1-|z|^\beta)^{3/2},  & {\rm if} |z|<1\\
 0,  & {\rm if} |z| \geq 1
\end{array} 
 \right.
 \label{eq12}
\end{equation}
where $z=\zeta/C_1,$ and $C_1$ is the eclipse half-width $(=D/200$, if using the eclipse full width in per cent, as defined in the GCVS \cite{samus} and required for the classification). 

The second-order trigonometrical polynomial is typically sufficient to describe the effects of reflection, ellipticity and asymmetry (O'Connell effect), and the $H-$ functions describe the shapes of the minima, with a parameter $\beta,$ which is generally different for the primary $(\beta_1=C_2$) and secondary  $(\beta_2=C_3$) minima. Generally, there may be a shift $\phi_0=C_4.$. Phenomenological modeling of multi-color observations of a newly discovered eclipsing binary 2MASS J18024395 + 4003309 = VSX J180243.9+400331 is presented in \cite{a2015}.

\begin{figure}[!h]
\centering
\epsfig{file = 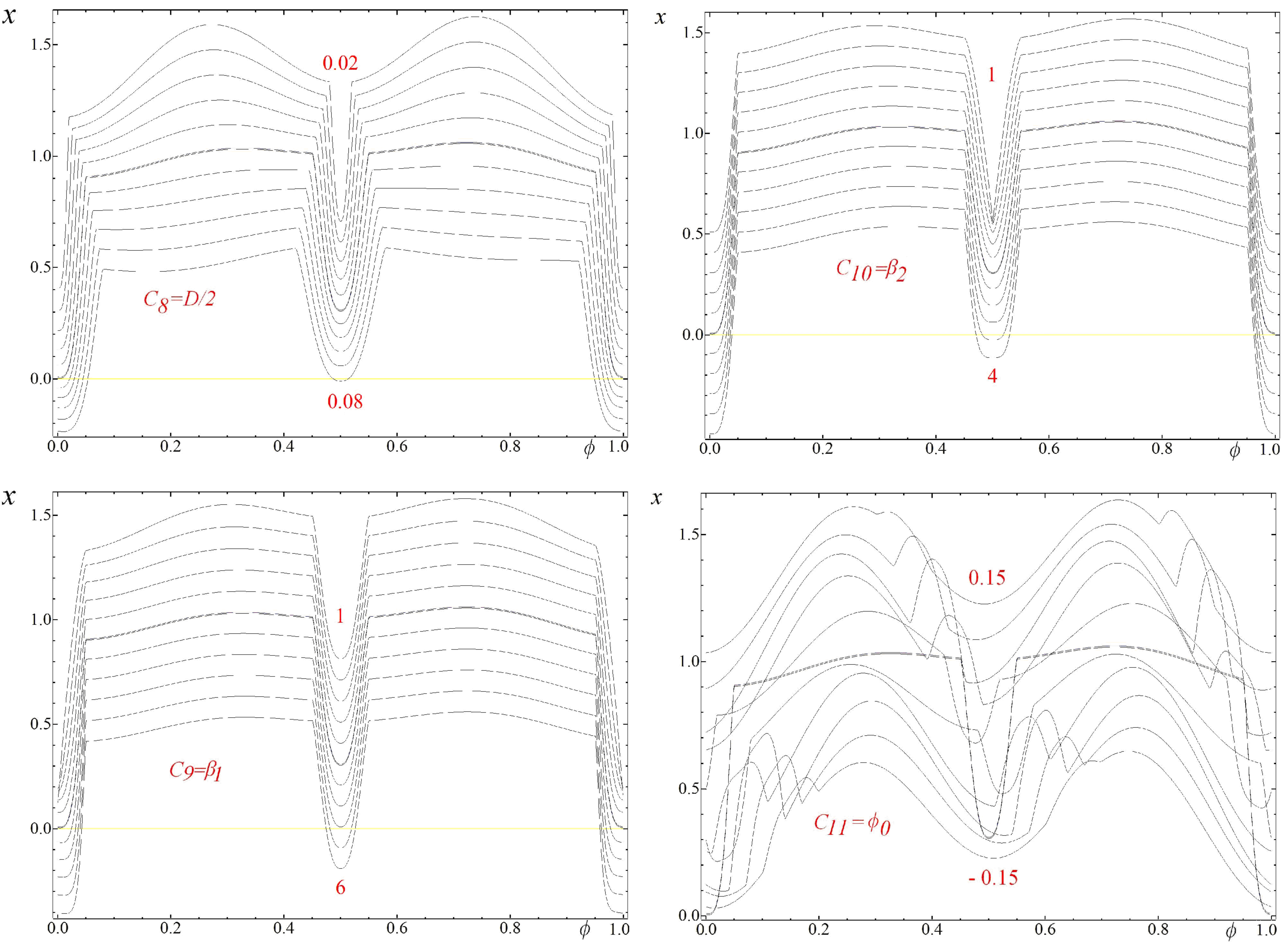,width = 1.0\linewidth}
\caption{Dependencies of the light curves (intensity vs. phase) on the parameters $C_8=D/2$ (left) and $C_9 = \beta_1$ (right). The relative shift in intensity between subsequent curves
is 0.1. The thick line shows a best fit curve.}\label{fig5}
\end{figure}

In Fig.~\ref{fig5}, we show dependencies of the best fit approximation with one parameter changing in a range, while other "non-linear" parameters $(C_1..C_4)$ are set to the best fit values, whereas the "linear" parameters $(G_1..G_7)$ are determined using the least squares subroutine.

The central thick line coincides with our artificial data, which were used for an illustration. It is clearly seen that the change of one of the "non-linear" parameters leads to changes in the "linear" parameters and thus the approximation.

In Fig.~\ref{fig6}, we show the "levels" - the lines of equal values of $\Phi_m$ at the two-parameter diagrams. They resemble deformed ellipses and show only a slight inclination close to the best fit point (marked by an arrow). The most drastic changes of the light curve are due to variations of the phase shift $C_4=\phi_0.$ There are only global minima of the function, except for the dependence with a phase shift $C_4.$ Such structure of the test function leads to the following algorithm of determination of the global minimum - at first, "brute force" determination of the minimum at a grid of values of $C_1..C_4$ with a further iterations using the differential corrections. However, if using the starting point at some middle point, the iterations may converge to a local minimum instead of the global one.

\section*{\sc conclusions}

The approximations with special pattern (also called "shape" or "profile") to fit the minima have much better quality of convergence of the smoothing curve with the data points. In this paper, we studied the dependence of the test function on four "non-linear parameters" $C_1..C_4,$ whereas the "linear" parameters $G_1..G_7$ are determined using the method of the least squares. The "NAV" ("New Algol Variable") algorithm is an effective tool presenting a good pattern for the minima, which may be improved by using an additional parameter, which describes its shape.

\begin{figure}[!h]
\centering
\epsfig{file = 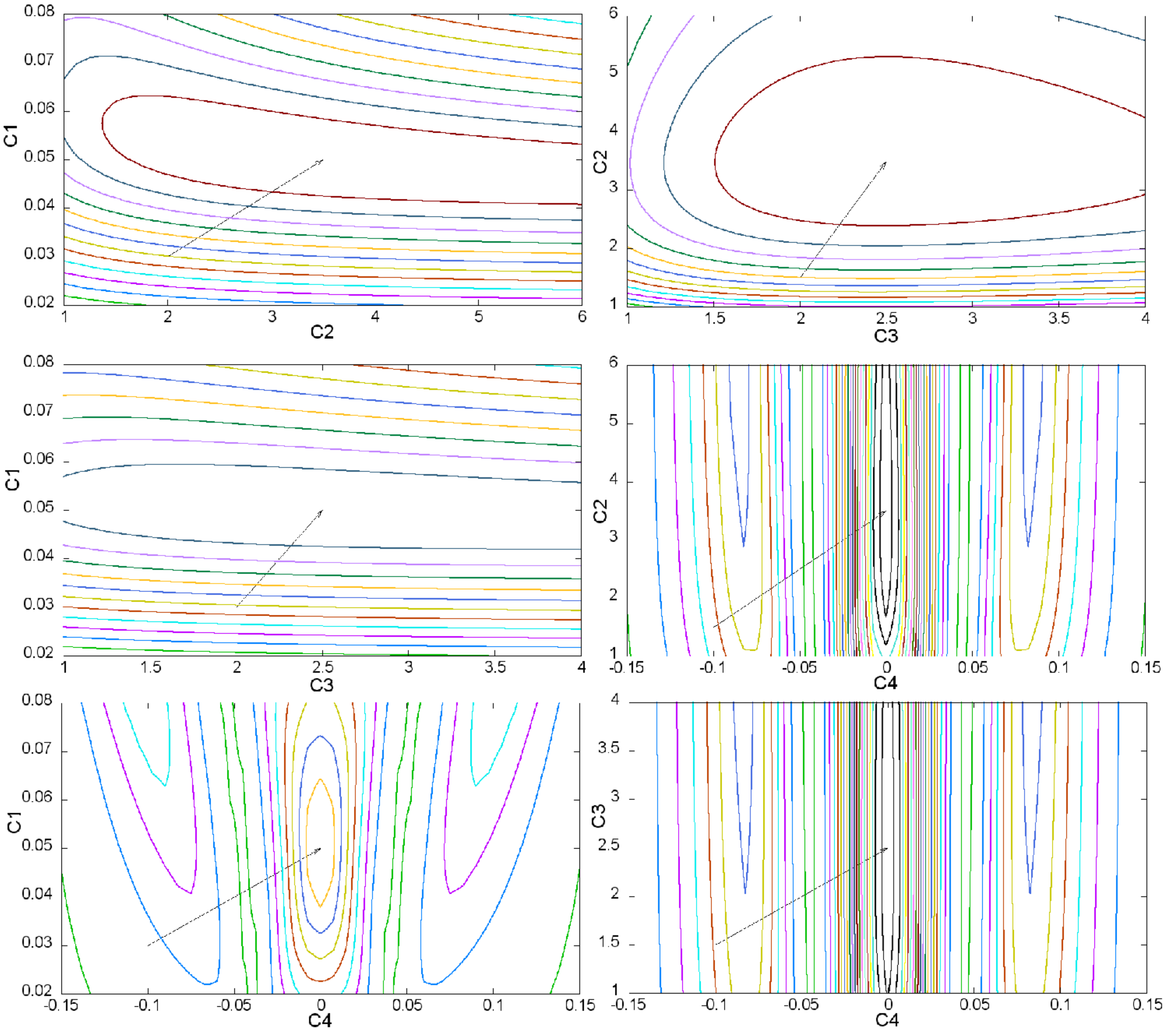,width = 1.0\linewidth}
\caption{Lines of equal levels of the test-function for different pairs of the parameters. Arrows show the best fit point.}\label{fig6}
\end{figure}


\end{document}